\documentclass[12pt]{article}
\usepackage{amsmath,amssymb,color}
\usepackage{cite,epsf} 
\usepackage{pstricks}
\usepackage{color}
\usepackage{hyperref}
\usepackage{tikz}     

\usetikzlibrary{
decorations.pathmorphing %library needed for e.g. decoration=snake
}

\usepackage{graphicx,array} 
\newcommand{\be}{\begin{equation}}
\newcommand{\ee}{\end{equation}}
\newcommand{\beq}{\begin{equation}}
\newcommand{\eeq}{\end{equation}}
\newcommand{\bea}{\begin{eqnarray}}
\newcommand{\eea}{\end{eqnarray}}

\newcommand{\ba}{\begin{eqnarray}}
\newcommand{\ea}{\end{eqnarray}}

\def\sin{\mbox{sin}}
\def\cos{\mbox{cos}}

\begin{document}

\begin{titlepage}
\vspace{10pt}
\hfill
{\large\bf HU-EP-23/01}
\vspace{20mm}
\begin{center}

{\Large\bf  Remarks on conformal invariants for\\[2mm] piecewise smooth curves and Wilson loops
}

\vspace{45pt}

{\large Harald Dorn 
{\footnote{dorn@physik.hu-berlin.de
 }}}
\\[15mm]
{\it\ Institut f\"ur Physik und IRIS Adlershof, 
Humboldt-Universit\"at zu Berlin,}\\
{\it Zum Gro{\ss}en Windkanal 6, D-12489 Berlin, Germany}\\[4mm]

\vspace{20pt}

\end{center}
\vspace{10pt}
\vspace{40pt}

\centerline{{\bf{Abstract}}}
\vspace*{5mm}
\noindent
This short note is some obvious mathematical  addendum to our papers on Wilson loops  on polygon-like contours with circular edges \cite{Dorn:2020meb,Dorn:2020vzj}. Using the technique of osculating spheres and circles we identify the conformal invariants characterising the kinks (cusps) of generic piecewise smooth curves in 3-dimensional space.

\vspace*{4mm}
\noindent

\vspace*{5mm}
\noindent
   
\end{titlepage}
\newpage

%\tableofcontents \newpage

\section{Introduction}
The metrical invariants of smooth curves in Euclidean D-dimensional space are length, curvature and $(D-2)$ torsion parameters as functions along the curve.
For piecewise smooth curves with cusps \footnote{We follow the language in the Wilson loop literature and  use the word cusp for a kink
with nonzero opening angle.}, in addition  each cusp is characterised by the Euler angles specifying the orthogonal  transformation needed to rotate the Frenet-Serret frames on both sides of a cusp to one another.

What concerns the issue of conformal invariants, there is for smooth curves a considerable amount of mathematical papers.  
In analogy to the metrical invariants length $s$, curvature $\kappa$ and torsion $\tau$ one  finds for 3-dimensional curves,  see e.g. \cite{Cairns},
conformal length $\omega$,\footnote{The prime denotes derivative with respect to $s$.}
\beq
d\omega~=~\sqrt{\nu}ds,~~~~~~\nu~=~\sqrt{(\kappa ')^2+\kappa ^2\tau ^2}~,\label{omega}
\eeq
conformal curvature $Q$
\beq
Q~=~\frac{4(\nu ''-\kappa ^2\nu)\nu -5(\nu ')^2}{8\nu ^3}~,\label{Q}
\eeq
and conformal torsion $T$
\beq
T~=~\frac{2(\kappa ')^2\tau+\kappa^2\tau ^3+\kappa\kappa'\tau '-\kappa\kappa''\tau}{\nu ^{\frac{5}{2}}}~.\label{T}
\eeq
Use of these invariants has been made in the physical literature  to characterise the boundary conditions for the treatment of minimal surfaces in AdS via Pohlmeyer reduction  \cite{He:2017cwd}.

But what concerns the extension to piecewise smooth curves, we did not find  any paper in the mathematical literature. Therefore, we decided to write up what one gets by straightforward application of one of the 
various techniques used for the smooth case: the kinematics of osculating spheres, see e.g. \cite{Fuster1,Fuster2,Langevin1,Langevin2} and refs. therein.

This note is organised as follows. In the next section we find formulas  expressing the conformal invariants for cusps, with legs
made of generic smooth pieces of curves, in terms of their metrical invariants.
Then in section 3 we consider the very special curves studied in our papers \cite{Dorn:2020meb,Dorn:2020vzj}, i.e. polygon-like curves with circular edges. These curves need a separate discussion, since along the circular edges all the conformal invariants \eqref{omega},\eqref{Q},\eqref{T} are zero or ill-defined.
\section{Conformal invariants of generic cusps}
In 3-dimensional space the osculating circle (sphere ) at a generic point $x$ on a smooth curve is a circle (sphere) having contact of
second (third) order with the curve at $x$. Since the order of contact is preserved under conformal transformations, osculating circles (spheres) at a given point of a curve are mapped to those for the image under a conformal map. Let us denote by $\vec t,\vec n,\vec b$ the unit tangent, normal and binormal vectors at $x$. Then the center $a$ of the osculating circle $S^1$ is given by 
\beq
a~=~x~+~\frac{1}{\kappa}~\vec n\label{circle}
\eeq
and the center $c$ of the osculating sphere $S^2$ by
\beq
c~=~x~+~\frac{1}{\kappa}~\vec n~-~\frac{\kappa '}{\tau\kappa ^2}~\vec b~.\label{sphere}
\eeq

We now turn to the case where $x$ is a point of discontinuity, i.e. the tip of a cusp. The cusp is then characterised by
two osculating circles $S^1_-,S^1_+$  and two osculating spheres $S^2_-,S^2_+$. The index $ " \pm "$ indicates
the limits one gets by approaching $x$ along  the both respective legs of the cusp.

Let us first count the number of conformal invariants we can expect for the cusp. There are 9 metrical invariants at hand:
$\kappa_{\pm},\kappa'_{\pm},\tau_{\pm}$ and the 3 Euler angles for the rotations from the Frenet frame $\{\vec t_-,\vec n_-,\vec b_-\}$ to $\{\vec t_+,\vec n_+,\vec b_+\}$. The difference of the numbers of parameters of the 3-dimensional conformal
and isometry group is equal to 4. This should result in $9-4=5$ conformal parameters. Now of course two of them are the corresponding limits of the differential of the conformal length \eqref{omega}. Hence there remain 3 conformal parameters to be attributed
genuinely to the cusp.

To  find explicit formulas for them, we start with the conformal invariants of pairs of spheres $(S^m,S^n)$ of the same or of different dimension. There is a rigorous mathematical  treatment for arbitrary dimensions in ref. \cite{Sulanke}. Applied to
our case it means that to each of the pair $(S^1_-,S^1_+),~(S^1_-,S^2_+),~(S^2_-,S^1_+),~(S^2_-,S^2_+)$ belongs just
one conformal parameter. To proceed, we first study these four invariants and will show afterwards, that only three of them
are independently.\\[2mm]
\underline{$(S^1_-,S^1_+)$}$~~~$
The tangents of the osculating circles agree with those of the curve. Therefore the related conformal invariant is 
\beq
A_{11}~=~\vec t_-\vec t_+~=~-\cos\ \alpha~. \label{A11}
\eeq
Here  $\alpha$ is the cusp angle (understood as the opening angle, i.e. $\alpha =\pi$ in the smooth case).\\[2mm]
\underline{$(S^2_-,S^2_+)$}$~~~$
Now the conformal invariant is given by the so-called inversive product, see e.g.\cite{Fuster1,Sulanke}
\beq
A_{22}~=~\frac{R_-^2+R_+^2-(c_--c_+)^2}{2R_-R_+}~=~\frac{(c_--x)(c_+-x)}{R_-R_+}~.\label{A22}
\eeq
$R_-$ and $R_+$ are the radii of $S^2_-$ and $S^2_+$. Obviously  $A_{22}  $ is   the cosine of the angle between
the vectors pointing from $x$ to the centers of the two osculating spheres. Strictly speaking, only its absolute value is invariant, since $A_{22}  $ changes sign under those special conformal transformations for which the preimage of infinity is situated inside just one of the spheres. The same comment applies to $A_{12}$ and $A_{21}$ below.\\[2mm]
\underline{$(S^2_-,S^1_+)$ and $(S^1_-,S^2_+)$}$~~~$For a sphere and an intersecting circle the conformal invariant is  the scalar product of the unit vector
pointing from $x$ to the center of the sphere with the unit tangent of the circle
\beq
A_{12}~=~\frac{\vec t_-(c_+-x)}{R_+},~~~
A_{21}~=~\frac{(c_--x)\ \vec t_+}{R_-}~.\label{A21}
\eeq

 Using \eqref{circle},\eqref{sphere} for both legs of the cusp we get from \eqref{A22} and \eqref{A21}
\beq
A_{22}~=~\frac{\kappa_-\tau_-\kappa_+\tau_+\ \vec n_-\vec n_++\kappa_-'\kappa_+'\ \vec b_-\vec b_+-\kappa_-\tau_-\kappa_+'\ \vec n_-\vec b_+-\kappa_+\tau_+\kappa_-'\ \vec b_-\vec n_+}{\sqrt{\kappa_-^2\tau_-^2+\kappa _-'^2}~\sqrt{\kappa_+^2\tau_+^2+\kappa _+'^2}}~,
\eeq
\bea
A_{12}&=&\frac{\kappa_+\tau_+\ \vec t_-\vec n_+-\kappa_+'\ \vec t_-\vec b_+}{\sqrt{\kappa_+^2\tau_+^2+\kappa _+'^2}}~,\\
A_{21}&=&\frac{\kappa_-\tau_-\ \vec t_+\vec n_- -\kappa_-'\ \vec t_+\vec b_-}{\sqrt{\kappa_-^2\tau_-^2+\kappa _-'^2}}~.
\eea
All the scalar products in the above formulas can be expressed in terms of three Euler angles $\varphi,\vartheta,\psi$ needed to rotate the Frenet frame $\{n_-,b_-,t_-\}$ to $\{n_+,b_+,t_+\}$, i.e.
\beq
\vec t_-\vec t_+=\cos\vartheta,~~~\vec t_-\vec b_+=\cos\varphi\, \sin\vartheta,~~~\vec b_-\vec t_+=-\cos\psi\,\sin\vartheta~.
\eeq

Then we get
\beq
A_{11}~=~\vec t_-\vec t_+~=~\cos\vartheta~,~~~~\mbox{i.e.} ~\vartheta=\pi-\alpha~,
\eeq
\bea
A_{22}&=&~\frac{1}{\sqrt{\kappa_-^2\tau_-^2+\kappa _-'^2}~\sqrt{\kappa_+^2\tau_+^2+\kappa _+'^2}}\Big ( \kappa_-\tau_-\kappa_+\tau_+\ (\cos\varphi\cos\psi-\sin\varphi\sin\psi\cos\vartheta) \nonumber\\
&&+\kappa_-'\kappa_+'(\cos\varphi\cos\psi\cos\vartheta-\sin\varphi\sin\psi)+\kappa_-\tau_-\kappa_+'\ (\sin\varphi\cos\psi+\cos\varphi\sin\psi\cos\vartheta)\nonumber\\
&&-\kappa_+\tau_+\kappa_-'\ (\cos\varphi\sin\psi+\sin\varphi\cos\psi\cos\vartheta)\Big )~,
\eea
\bea
A_{12}&=&\frac{\sin\vartheta}{\sqrt{\kappa_+^2\tau_+^2+\kappa _+'^2}}\ \big (\kappa_+\tau_+\ \sin\varphi-\kappa_+'\ \cos\varphi\big )~,\\
A_{21}&=&\frac{\sin\vartheta}{\sqrt{\kappa_-^2\tau_-^2+\kappa _-'^2}}\ \big (\kappa_-\tau_-\ \sin\psi +\kappa_-'\ \cos\psi\big )~.
\eea
The part of $A_{22}$ containing the factor $\cos\vartheta$ is related to the product of $A_{12}$ and $A_{21}$ in an obvious manner. With
a  bit more careful  inspection one gets for the whole  $A_{22}$
\beq
A_{22}(\varphi,\vartheta,\psi)~=~\frac{1}{\sin^2\vartheta}\Big ( A_{12}(\varphi+\frac{\pi}{2})\ A_{21}(\psi+\frac{\pi}{2})-\cos\vartheta \ A_{12}(\varphi)A_{21}(\psi)\Big )\label{A22A12}~.
\eeq
Now $A_{12}$ and $A_{21}$ for arguments shifted by $\frac{\pi}{2}$ are not independent, but related by
\beq
\big (A_{12}(\varphi)\big )^2+\big (A_{12}(\varphi+\frac{\pi}{2})\big )^2~=~(A_{21}(\psi)\big )^2+\big (A_{21}(\psi+\frac{\pi}{2})\big )^2~=~\sin^2\vartheta\label{A22A12b}~.
\eeq
This means that a complete set of independent conformal invariants attributed to the cusp is given by  the three parameters  \footnote{Remember  $\varphi,\vartheta,\psi$
Euler angles,  $\alpha =\pi-\vartheta$ opening angle of the cusp.}
\beq
\boxed{\alpha~, ~~~ B_{12}= \frac{\kappa_+\tau_+\ \sin\varphi-\kappa_+'\ \cos\varphi}{\sqrt{\kappa_+^2\tau_+^2+\kappa _+'^2}}~,~~~B_{21}=\frac{\kappa_-\tau_-\ \sin\psi+\kappa_-'\ \cos\psi}{\sqrt{\kappa_-^2\tau_-^2+\kappa _-'^2}}~.}
\eeq 

Let us add a warning. Inserting by brute force $\varphi=\psi=0$ into $B_{12}$ and $B_{21}$ one could reach the wrong conclusion that
$\frac{\kappa'}{\sqrt{\kappa^2\tau^2+\kappa '^2}}$ could be an invariant for smooth curves. But since putting $\varphi$ or $\psi$ to zero
is not a conformal invariant statement, this conclusion is not allowed and also straightforwardly proven to be wrong.\\

There is still another approach to the invariants $B_{12}$ and $B_{21}$.  With conformal transformations each smooth curve can be  mapped 
to a normal form with respect to a chosen point on the curve. Then  the image  of that point is at  the origin of some Cartesian coordinates 
and the curve gets the form  
\cite{Cairns,He:2017cwd,Langevin1}
\bea
x_1(u)&=&u~,\nonumber\\
x_2(u)&=&\frac{u^3}{3!}~+~(2Q-T^2)\,\frac{u^5}{5!}~+~{\cal O}(u^6)~,\nonumber\\
x_3(u)&=&T\,\frac{u^4}{4!}~+~\frac{1}{\sqrt{\nu}}\,\frac{dT}{ds}\,\frac{u^5}{5!}~+~{\cal O}(u^6)~,\label{normal}
\eea
with  $Q$ and $T$ the conformal curvature and torsion at the point under consi\-deration.  The related metrical invariants at $x(0)$ are
\beq
\kappa ^{(N)}~=~0~,~~~\kappa'^{(N)}~=~1~,~~~\tau^{(N)}~=~\frac{T}{2}~.\label{Normal}
\eeq

What is now the normal form of a curve with a cusp just with respect to the tip of the cusp ? There are two. The first is obtained by mapping the piece  before the cusp to the form \eqref{normal}, with $Q,\,T$ replaced by $Q_-,\,T_-$. The second normal form  one gets by mapping the piece after the cusp to the form  \eqref{normal}, with $Q_+,\,T_+$. Of course in both cases the respective other leg of the cusp is not of the form \eqref{normal}.  Calculating now  $B_{21}$ in the first
normal form and $B_{12}$ in the second one, with the use of \eqref{Normal} for the $\pm$ versions, we get
\bea
B_{21}&=&\cos\,\psi^{(N1)}~,\nonumber\\
B_{12}&=&-\cos\,\varphi^{(N2)}~.
\eea
$\psi^{(N1)}$ and $\varphi^{(N2)}$ are Euler angles in the first and second normal form, respectively.\\

We continue with some casual comment on the conformal invariants in 4-dimensio\-nal space. To fix all the osculating spheres from $S^1$
to $S^3$ at a smooth point one needs $\kappa,\kappa ',\kappa'',\tau_1,\tau_1',\tau_2$. This for the limits from both sides of the cusp, together
with 6 Euler angles, needed in 4D for the rotation of the Frenet frame, gives 18 metrical parameters. The difference of the number of parameters
oft the conformal and isometry group is now 5, hence we tentatively reach 13 conformal parameters. Now on both sides the limits of the differential
of the conformal length and the first conformal torsion\footnote{In 3D the conformal torsion \eqref{T} cannot be build out of  the metrical parameters needed to fix all the osculating spheres at the cusp. But in 4D the osculating $S^3$ inherits all the information.} are not related to the cusp. Hence we can expect $13-4=9$ conformal parameters to be attributed genuinely to the cusp.

On the other side, among the nine pairs of osculating spheres \\$(S^i_-,S^j_+),\ i,j=1,2,3$, the pairs $(S^2_-,S^2_+),\ (S^2_-,S^1_+),\ (S^1_-,S^2_+)$ have two invariants and all other only one\cite{Sulanke}. This yields $12$ conformal parameters. We keep it as an open question, whether there are
indeed three relations of the 3D type \eqref{A22A12},\eqref{A22A12b} to reach the minimal number of 9 independent parameters seen in the previous paragraph.

We close this section with a comment on the cusp anomalous dimension for Wilson loops. It is generally believed to depend on the cusp
angle $\alpha$ only, and therefore calculations have been done using straight edges. While in field theoretic perturbation theory this
can be justified by power counting in the corresponding Feynman integrals, a rigorous proof for the generic situation at strong coupling is still lacking. We have presented a proof for the planar case with generically curved edges in \cite{Dorn:2015bfa}. For full generality in 3D, a proof 
of its independence from  $B_{12}$ and $B_{21}$ is still lacking.
  
\section{Comments on Wilson loops  on polygon-like \\curves with circular edges}
The polygon-like curves with circular edges, whose related Wilson loops have been  studied in our papers \cite{Dorn:2020meb,Dorn:2020vzj},
are not covered by the setting for generic curves as presented in the previous section. Along their edges one has constant curvature $\kappa$ and zero
torsion $\tau$, resulting in zero conformal length \eqref{omega} and undefined conformal curvature and torsion. In mathematical language these edges are conformal vertices.\footnote{See e.g. \cite{Cairns,Fuster1,Fuster2}.} 

Although such  single edges carry no conformal data, their combination to a polygon does\footnote{ In a sense one could
call it a conformal vertex with internal conformal substructure.} It has still no extension in the sense of conformal length, but there are of course
the cusp angles at each cusp of the polygon and for more than 3 cusp points the corresponding cross ratios. In addition, like generic curves, these
polygons can wind themselves out of a plane and exhibit torsion.   In our paper \cite{Dorn:2020meb} we have this issue parameterised by the introduction of torsion angles $\beta_j$, defined at a given cusp point $x_j$ by
\beq
\beta_j~=~\measuredangle (\{x_{j},x_{j+1}\},cc_j)~,\label{beta}
\eeq
 where $\{x_{j},x_{j+1}\} $ denotes the circular edge between $x_j$ and $x_{j+1}$ and $cc_j$ the circle fixed by the three cusp points $x_{j-1},x_j,x_{j+1}$.

However, in contrast to  $B_{12}$ and $B_{21}$ for cusps of generic curves, these torsion angles are not attributed to the local properties at the corresponding cusp. This is simply seen by changing in \eqref{beta}
 the neighbouring cusp point $x_{j+1}$ along the circle of which the edge $\{x_{j},x_{j+1}\} $ is a part . Then the tangent of the edge at $x_j$ remains the same,  but the circle $cc_j$ and its tangent at $x_j$ changes. By this manipulation $\beta_j$ changes, although the local situation at the cusp at $x_j$ remains the same as before.

We end with a remark on some setting for Wilson loops on piecewise smooth curves  intermediate between 
those of  full generality considered in section 2 and the polygons with circular edges in \cite{Dorn:2020meb}. In conformal invariant gauge field theories as in e.g. ${\cal N}=4$ SYM  there will be an anomalous conformal Ward identity of the type derived in \cite{Dorn:2020meb},
imposing for the Wilson loop the structure  of a conformally covariant factor depending on the distances of the
tips of the cusps times a conformally invariant remainder factor.  In the generic case this remainder factor is a 
function of the cross ratios formed out of cusp points and the conformal cusp parameters identified in section 2, but in addition also a functional of  the conformal invariants as functions along the edges. For the case of polygons with circular edges the remainder is a function of only a finite number of conformal parameters. 

To have a curve characterised  by a finite number of conformal parameters but nevertheless having nonzero
conformal length, one could consider polygons with edges which are pieces of curves with constant conformal curvature and torsion.
Such curves have been classified in\cite{Sulanke2}. Among them are loxodromes on rotational surfaces.\\[20 mm]
{\bf Acknowledgement}\\[2mm]
I thank the Quantum Field and String Theory Group at Humboldt University for kind hospitality.
%%%%%%%%%%%%%%%%%%%%%%%%%%%%%%%% 

\end{document}